\documentclass[10pt]{article} 
\usepackage{amsfonts,amssymb,slashed,makeidx,latexsym,setspace}
\usepackage{graphicx,graphics,floatflt,amssymb,epsf,rotate,subfigure} 
\usepackage{times,cite,color}
\usepackage{multirow}
\usepackage{hyperref}
\begin{document}

\begin{flushright}
SINP/TNP/2012/13, DO-TH 12/29
\end{flushright}

\begin{center}
{\Large \bf Pushing the SUSY Higgs mass towards 125 GeV with a color
  adjoint} \\
\vspace*{1cm} \renewcommand{\thefootnote}{\fnsymbol{footnote}} { {\sf
    Gautam Bhattacharyya${}^{1,2}$} and {\sf Tirtha Sankar Ray${}^{3,4}$}}
\\
\vspace{10pt} {\small ${}^{1)}$ {\em Saha Institute of Nuclear
    Physics, 1/AF Bidhan Nagar, Kolkata 700064, India} \\ 
${}^{2)}$ {\em Fakult\"at f\"ur Physik,
Technische Universit\"at Dortmund, 44221 Dortmund,
Germany} \\ 
${}^{3)}$ {\em Institut de Physique Th\'eorique, CEA-Saclay, F-91191
    Gif-sur-Yvette Cedex, France} \\
${}^{4)}$ {\em ARC Centre of Excellence for Particle Physics at the Terascale,
School of Physics, \\University of Melbourne, 
Victoria 3010, Australia}} 
\normalsize
\end{center}

\begin{abstract}
We show that inclusion of a TeV scale chiral superfield transforming
in the adjoint representation of the color SU(3) to the MSSM particle
content modifies the renormalization group running of some parameters
in such a way that a 125 GeV mass of the light Higgs boson is
accommodated more comfortably than in cMSSM / mSUGRA. Put differently,
the introduction of a color adjoint TeV scale superfield helps
resurrecting lighter choices for the stop and gluino which are
otherwise disfavored in cMSSM / mSUGRA.
\end{abstract}

\setcounter{footnote}{0}
\renewcommand{\thefootnote}{\arabic{footnote}}

\noindent {\bf Introduction:}~ The discovery of the Higgs boson, or
more appropriately, a Higgs-like boson at the CERN Large Hadron
Collider (LHC) \cite{:2012gk,:2012gu} has pushed one of the most
advertised class of supersymmetric (SUSY) models, namely, the
constrained minimal supersymmetric standard model (cMSSM), or
equivalently, minimal supergravity (mSUGRA), into an uncomfortable
corner. The disappointment arising from the so far unsuccessful
attempts to underpin any supersymmetric relics via direct searches at
the LHC \cite{atlassusy,cmssusy} got further aggravated by the news
that the Higgs boson is as heavy as 125 GeV. This is so because the
third generation squark masses and the associated soft trilinear
scalar coupling, on which LHC cannot put as stringent direct
constraints as on their first two generation counterparts, are now
pushed to one to few TeV to ensure that the lightest Higgs mass
receives sufficient radiative enhancement.  It is in this context that
we write this short note. We propose that a simple augmentation of the
MSSM particle content with a chiral superfield transforming in the
adjoint representation of the SU(3) color group proves to be useful in
easing part of this difficulty and thus resurrecting some of the lost
parameter space. Put briefly, our scenario is the following: add a
colored chiral superfield whose scalar and fermionic components behave
like a scalar gluon and a gluino, respectively, with no other
non-vanishing gauge quantum numbers, each weighing around a TeV.
Their presence would modify the renormalization group (RG) running of
the QCD coupling $g_3$ (more specifically, would add positive
contributions to its beta function), which would in turn feed into the
running of the top Yukawa coupling $y_t$, and the trilinear scalar
coupling $A_t$ entering the stop mixing matrix. These modifications
give us a few {\em territorial} advantages over cMSSM: ($i$) the
gluino and the lighter stop can be lighter than what they should weigh
in cMSSM for generating the 125 GeV mass of the light Higgs; ($ii$)
the rather large (possibly maximal) stop-mixing, which facilitates
reaching out to 125 GeV mass of the Higgs, does not compel $|A_0|$ any
more to be as large as what cMSSM requires it to be.  The motivation
for adding an adjoint superfield may come from string theory, more
specifically, the intersecting D-brane models \cite{Cui:2006nw}.
However, such color adjoint fields in four dimensional context need
not be seen only as spies from extra dimension or string theory, they
may very well have more mundane ancestry. The main upshot of our
analysis is that the introduction of a color adjoint superfield at the
TeV scale improves the fine-tuning.

Other options for creating more room for accommodating a 125 GeV SUSY
Higgs also exist. Next-to-minimal supersymmetry (NMSSM) is already
known to possess an improved fine-tuning as its gauge singlet
superfield, coupled to the two Higgs doublets in the superpotential,
provides a tree level mass to the lightest CP-even Higgs
\cite{Drees:1988fc,Ellwanger:2009dp}.  A recent numerical study of the
trilinear scalar couplings $A_\lambda$ and $A_\kappa$ in the
conventional scale invariant version of NMSSM, however, shows that the
125 GeV mass of the Higgs boson is compatible only in some
well-separated islands of the parameter space \cite{Agashe:2012zq}.
Further reduction in fine-tuning in NMSSM has more recently achieved
by introducing extra matter descending from $E_6$ origin in a scenario
which also possesses a discrete $R$-symmetry solving the domain wall
problem and enhancing proton stability \cite{Hall:2012mx}.  A
bottom-up approach for addressing the fine-tuning problem, which goes
by the name of `natural SUSY', has also gained attention where the
third generation sfermions and the Higgsino are kept light, while the
rest of the superpartners are considered heavy
\cite{Espinosa:2012in,Papucci:2011wy,Brust:2011tb}. Additional matter
fields transforming as grand unified theory (GUT) multiplets have also
been employed to {\em better} realize the 125 GeV mass of the Higgs,
improving consistency with the muon $(g-2)$ measurement at the same
time \cite{Endo:2011mc, Moroi:2011aa}. Our method of {\em comfortably
  achieving} the 125 GeV mass of the Higgs relies on adding a color
adjoint state that does not directly couple to the Higgs sector but
its effect filters through to the Higgs mass only via modifications of
RG running of various couplings.  Adjoint representation states have
been employed for different purposes so far.  It was shown that such a
colored chiral superfield, appearing e.g. in the context of a
4-dimensional realization of $N=2$ supersymmetry in a 5-dimensional
theory \cite{ArkaniHamed:2001tb,Bhattacharyya:2010rm}, helps the
gluino acquire a large Dirac mass in a class of super-soft SUSY
breaking models \cite{Fox:2002bu} actually helps to improve the
fine-tuning \cite{Kribs:2012gx} of parameters. A very recent study
\cite{Perez:2012gg} aiming to improve the fine-tuning has employed the
24-plet SU(5)-adjoint superfield, the vacuum expectation value of
whose singlet component helps to enhance the tree level Higgs mass in
the NMSSM-style. As a result, the 125 GeV Higgs mass is reached with a
lighter stop and smaller mixing, the colored states being used to keep
the running of the top Yukawa coupling under control.

\begin{figure*}[htbp]
\begin{minipage}[]{0.44\textwidth}
\begin{center}
\includegraphics[width=0.7\textwidth,angle=270,keepaspectratio]
{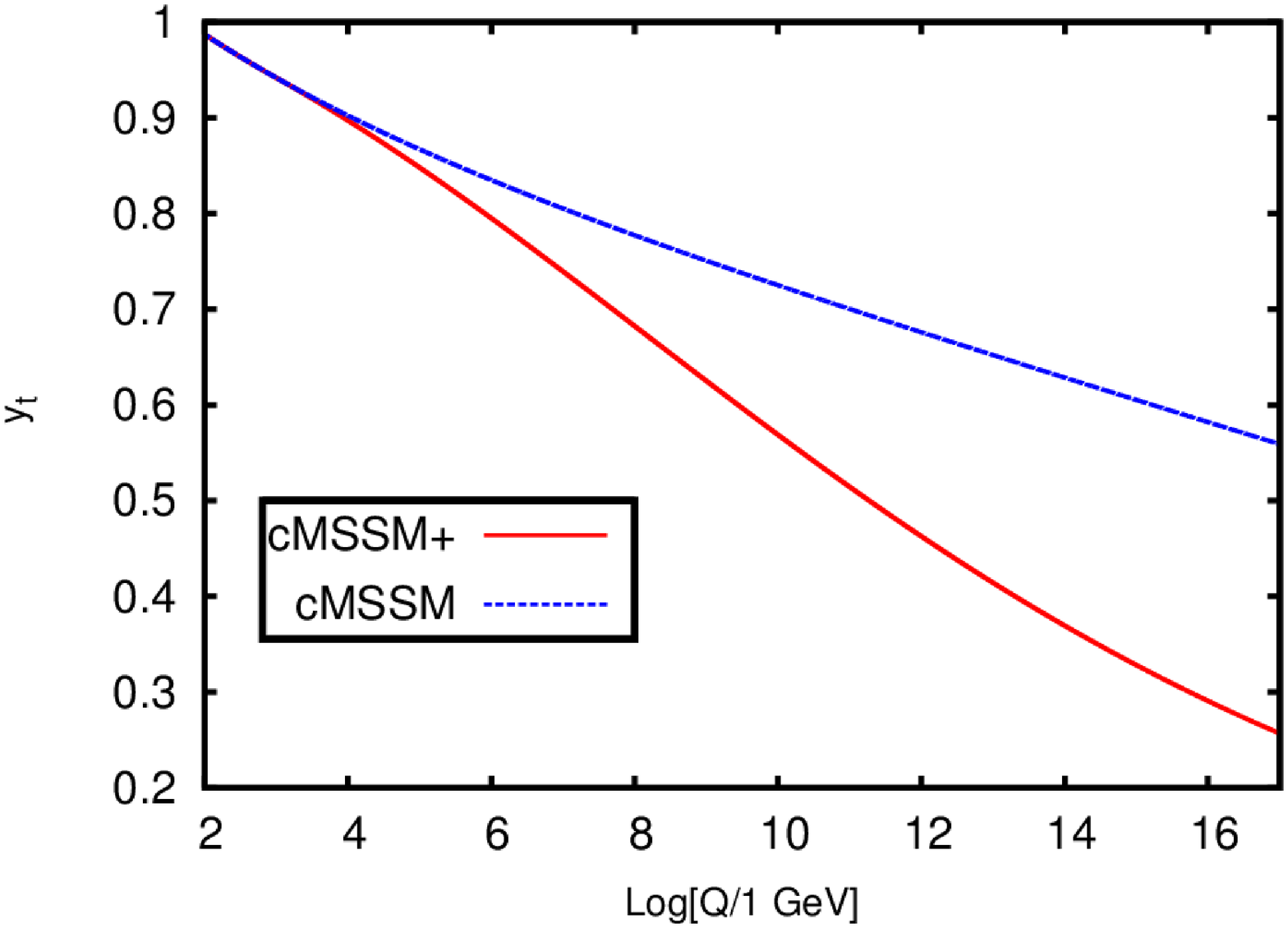}
\caption[]{\em \small RG running of the top Yukawa coupling in cMSSM+
  (red solid line) and in cMSSM (blue dashed line).}
\label{fig1}
\end{center}
\end{minipage}
\hspace{7mm}
\begin{minipage}[]{0.44\textwidth}
\begin{center}
\includegraphics[width=0.7\textwidth,angle=270,keepaspectratio]
{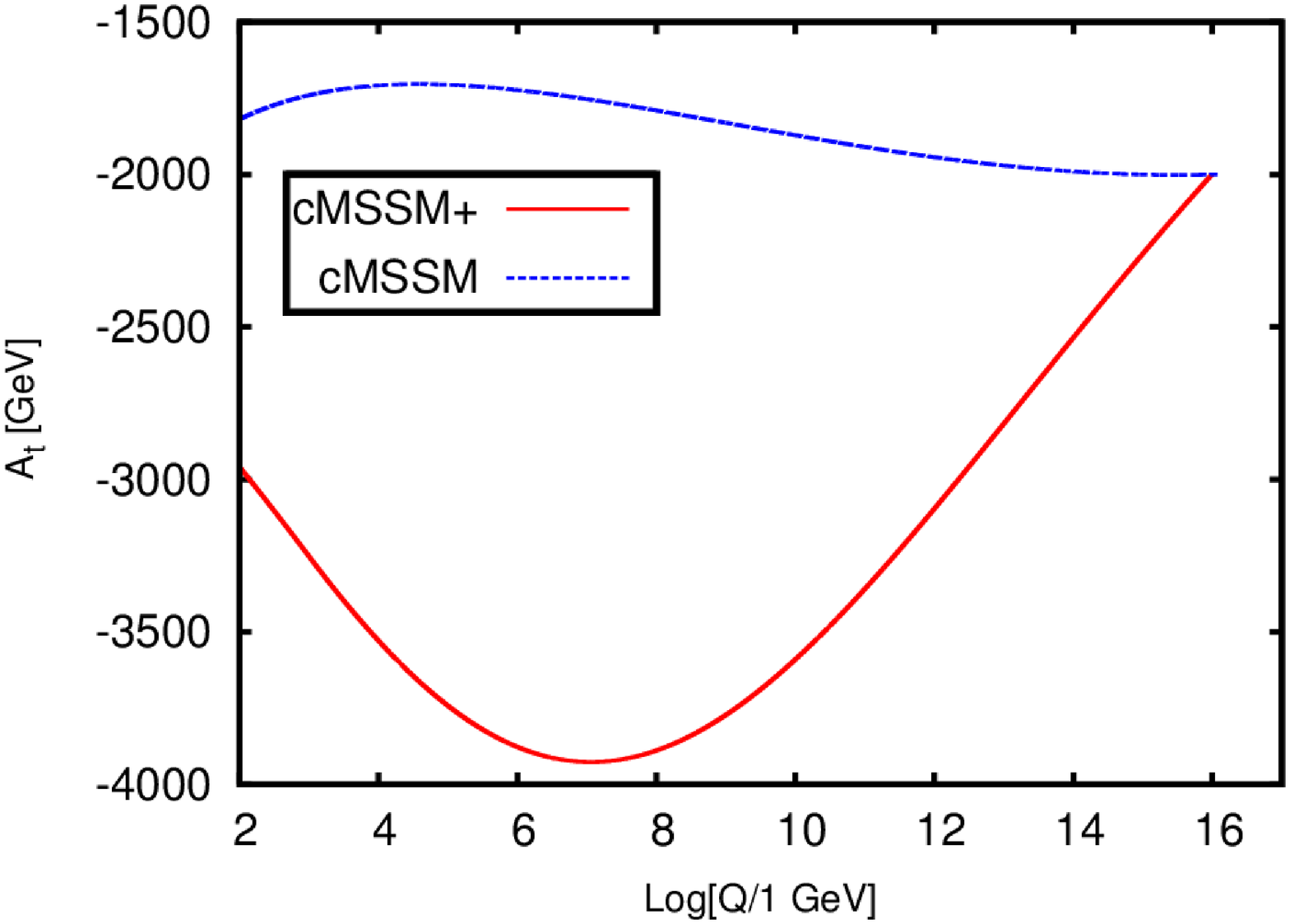}
\caption[]{\em \small RG running of the trilinear scalar parameter
  $A_t$ in cMSSM+ (red solid line) and in cMSSM (blue dashed
  line). Here, $M_{1/2} = 500$ GeV and $\tan\beta = 10$.}
\label{fig2}
\end{center}
\end{minipage}
\end{figure*}

\noindent {\bf Formalism:}~ We refer to the present scenario as
`cMSSM+' for repeated use in the subsequent text, which implies the
MSSM particle content plus a single ${\rm SU(3)_C}$ adjoint chiral
superfield.  We now demonstrate how its introduction induces a
drastic modification to RG evolution of several parameters. All we
need to calculate are the contributions of the fermion and scalar
components of the adjoint superfield to the QCD gauge beta
function. These are given by
\begin{eqnarray}
\label{b3}
\Delta b_3^f = \frac{4}{3} \cdot 3 \cdot \frac{1}{2} = 2 \, , ~~~{\rm and}~~~
\Delta b_3^s = \frac{1}{3} \cdot 3 = 1 \, ,  
\end{eqnarray}
where the factor 3 represents color, and the factor (1/2) in the
fermionic contribution comes from its Majorana nature. Hence $\Delta
b_3 = \Delta b_3^f + \Delta b_3^s = 3$.  We assume that the new
fermion and the scalar weigh around a TeV. So, as soon as this energy
is crossed, the new states are sparked into life, and the above
increment in the beta function changes the slope of the running of
$g_3$ keeping it flat at its weak scale value all along (up to one
loop precision). This constitutes the primary effect and the rest is
simply its consequence, as we explain now step by step.  We recall
that the gauge beta functions are given by ($t = \ln Q/(1~{\rm
  GeV})$),
\begin{eqnarray}
\label{beta-gauge}
\beta_{g_a} \equiv {d\over dt} g_a =  {b_a\over 16\pi^2} g_a^3, 
\end{eqnarray}
where $(b_1, b_2, b_3) = (33/5,\> 1,\> -3)$ for MSSM at one loop
\cite{Martin:1997ns}. Since, only $b_3$ receives an increment in
cMSSM+, as shown in Eq.~(\ref{b3}), the cMSSM+ set reads: $(b_1, b_2,
b_3) = (33/5,\> 1,\> 0)$. For our purpose, one loop estimate of beta
functions is enough\footnote{Two loop RG evolution in Dirac Gaugino
  context has been discussed in \cite{Goodsell:2012fm}.}. Since in
cMSSM+ $g_3$ hardly runs beyond the TeV scale, the gluino mass also
remains stationary at the leading order.  Admittedly, gauge couplings
do not unify in this model since only the slope of $g_3$ running is
modified, although the value of $g_3$ remains perturbative all the way
up to the high scale\footnote{Additional chiral multiplets suitably
  charged under ${\rm SU(2)_L}$ and ${\rm U(1)_Y}$ may be added to
  ensure that the RG curves of $\alpha_2$ and $\alpha_1$ are also bent
  appropriately to reinstate gauge coupling unification at a value
  higher (still perturbative) than in MSSM. But this is not our main
  focus and we do not pursue the unification issue any further. For a
  discussion on gauge coupling unification in F-theory GUT models with
  Dirac gauginos, see \cite{Davies:2012vu}.}.  In our subsequent
numerical discussions on cMSSM+ we treat $M_G = 2 \cdot 10^{16}$ GeV
as a high scale reference point (for comparison of various running
{\em vis-\`a-vis} MSSM), and assume that the common gaugino
and scalar supersymmetry breaking mass parameters, $M_{1/2}$ and
$m_0$, respectively, refer to that point. We now look at the RG
running of the top Yukawa coupling, where for illustration we display
only the dominant terms:
\begin{eqnarray}
\label{beta-top}
 \beta_{y_t} \equiv {d\over dt} y_t \simeq  {y_t \over 16 \pi^2} 
\Bigl [6 y_t^* y_t - {16\over 3} g_3^2\Bigr ].
\end{eqnarray}
Since $g_3$ stays at the large weak scale value even at the high
scale, the RG trajectory of the top Yukawa coupling is bent to lower
values compared to MSSM at the high scale -- see Fig.~\ref{fig1}. This
will help us understand the evolution pattern of the trilinear
coupling $A_t$. Again, we display the dominant terms for providing
intuition:
\begin{eqnarray}
\label{beta-at}
16\pi^2 {d\over dt} A_t \simeq A_t \Bigl [18 y_t^* y_t - {16\over 3} g_3^2 \Bigr]
+ {32\over 3} y_t g_3^2 M_3 \, . 
\end{eqnarray}
An interplay of Eqs.~(\ref{beta-gauge}), (\ref{beta-top}) and
(\ref{beta-at}) provides the insight that starting from a given
negative high scale value $A_0$, the weak scale value $A_t$ is more
negative in cMSSM+ compared to cMSSM -- see Fig.~\ref{fig2} (drawn for
$M_{1/2} = 500$ GeV and $\tan\beta = 10$).  It is now known that a
negative $A_t$ of larger magnitude is more helpful for reaching out to
125 GeV mass of the Higgs (see e.g.
\cite{Brummer:2012ns,Ghosh:2012dh} for recent studies). This
transpires from
\begin{eqnarray}
\label{mh}
 m_h^2 = M_Z^2 \cos^2 2\beta +
 \frac{3}{4\pi^2}\frac{m_t^4}{v^2}\left[\log\frac{M_S^2}{m_t^2} +
 \frac{X_t^2}{M_S^2} \left(1-\frac{X_t^2}{12M_S^2}\right)\right] \, , 
\end{eqnarray}
where $v = \sqrt{v_u^2+v_d^2} = 174$ GeV, $\tan\beta = v_u/v_d$, $X_t
\equiv A_t-\mu\cot\beta$, and $M_S \equiv
\sqrt{m_{\tilde{t_1}}m_{\tilde{t_2}}}$ is the geometric mean of the
two stop masses. Although Eq.~(\ref{mh}) does not care about the sign
of $A_t$, but given the slope of its RG trajectory a large negative
value of $A_t$ is easier achieved than a positive value of the same
magnitude.

\begin{figure*}[htbp]
\begin{minipage}[]{0.44\textwidth}
\begin{center}
\includegraphics[width=0.7\textwidth,angle=270,keepaspectratio]
{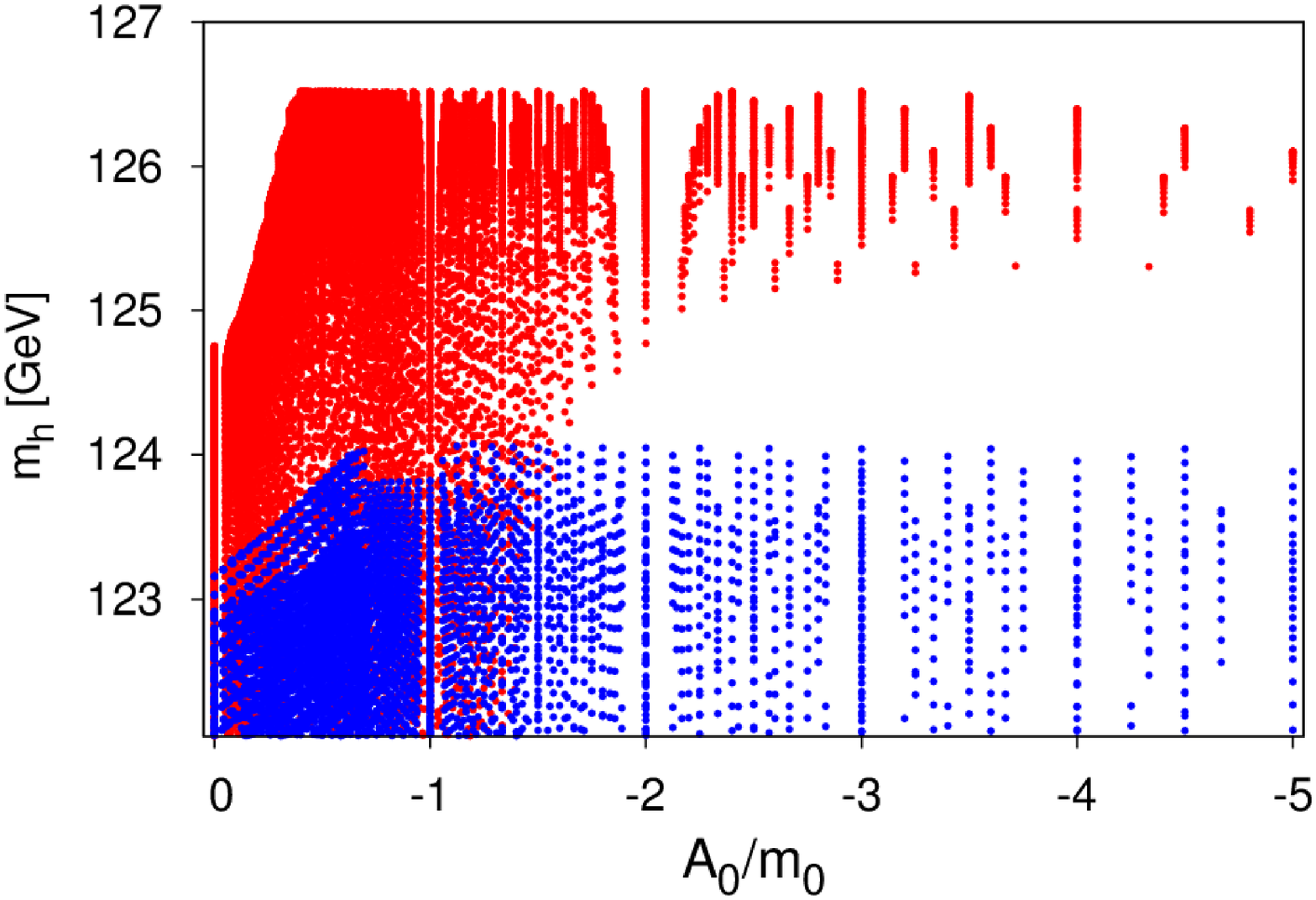}
\caption[]{\em \small Scatter plot of the Higgs mass where the high
  scale parameters are fixed at $M_G = 2 \cdot 10^{16}$ GeV (red
  lighter points in the top patch for cMSSM+, blue darker points in
  the bottom patch for cMSSM). A systematic enhancement in the Higgs
  mass by about 3 GeV is noticed for cMSSM+ as compared to cMSSM for
  the same choice of $A_0/m_0$.}
\label{fig3}
\end{center}
\end{minipage}
\hspace{7mm}
\begin{minipage}[]{0.44\textwidth}
\begin{center}
\includegraphics[width=0.7\textwidth,angle=270,keepaspectratio]
{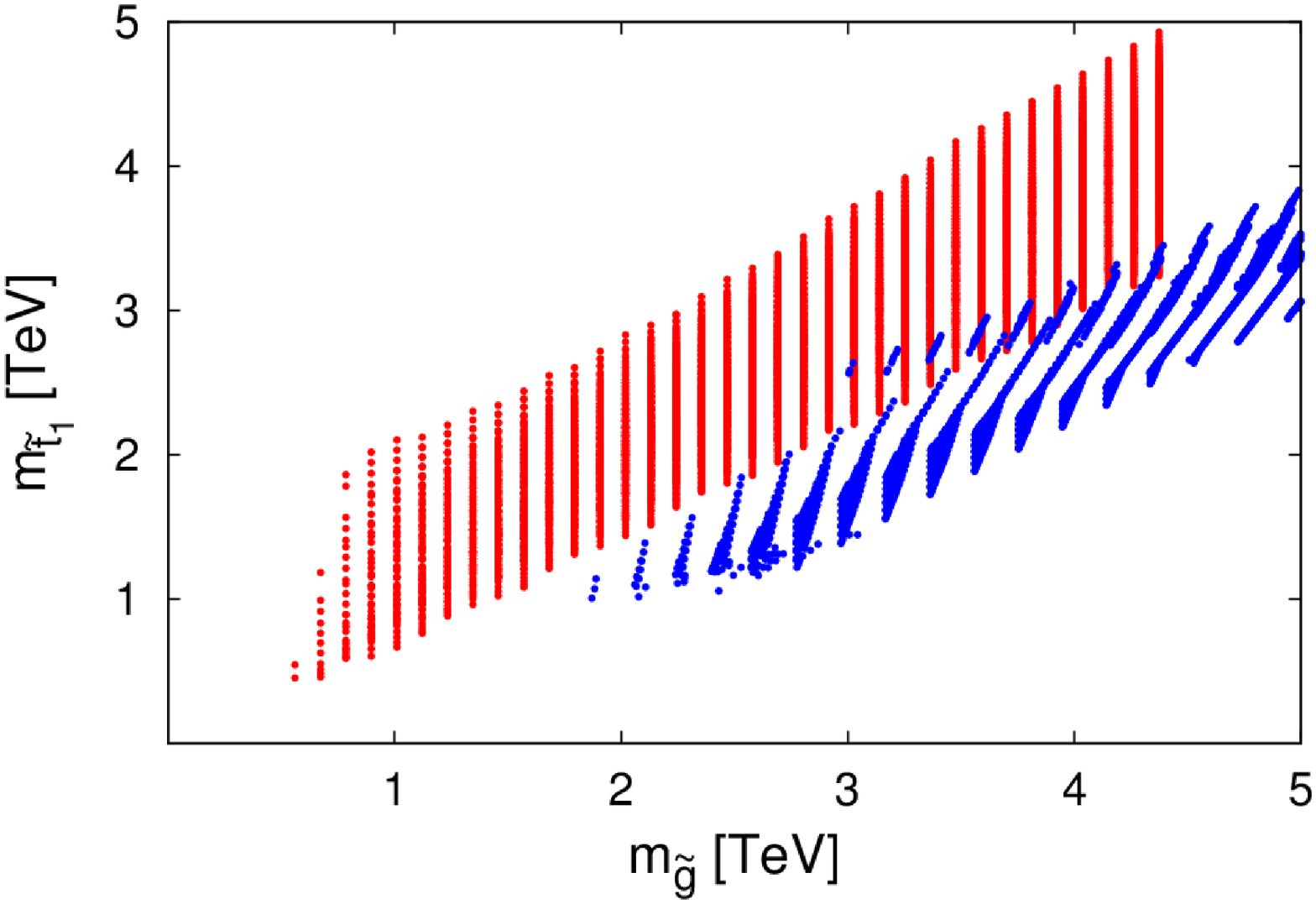}
\caption[]{\em \small The shaded regions correspond to $m_h$ between
  123 and 127 GeV. A significant shift towards a lighter stop and
  gluino is observed for cMSSM+ (red lighter points) as compared to
  cMSSM (blue darker points).}
\label{fig4}
\end{center}
\end{minipage}
\end{figure*}

\noindent {\bf Results and other implications:}~ In Fig.~\ref{fig3} we
show the scatter plot of the Higgs mass for different choices of the
ratio of high scale parameters $A_0$ and $m_0$ in cMSSM+ (red points)
and in cMSSM (blue points). We scan over the following ranges:
$m_0=[0,2]$ TeV, $M_{1/2}=[0,2]$ TeV and $A_0=[0,-2]$ TeV, keeping
$\tan\beta=10$ and $\mu>0$ (preferred by $(g-2)$ of muon).  What is
significant is that for the above parameter choices, especially, the
magnitude of negative $A_0$ not exceeding 2 TeV, cMSSM struggles to
give the Higgs a mass of 125 GeV \cite{Brummer:2012ns, Ghosh:2012dh,
  Cao:2011sn}, but for the same ranges of parameters cMSSM+ offers a 3
to 4 GeV enhancement to the Higgs mass which is enough to bring it
into consistency with the CMS and ATLAS measurement.  This is simply a
consequence of a more negative value of $A_t$ that is attainable in
cMSSM+ compared to what is possible in cMSSM, i.e. $|A_t~({\rm
  cMSSM+})| > |A_t~({\rm cMSSM})|$, starting from a given (negative)
$A_0$ at the high scale. This happens because the running of $A_t$ has
a steeper slope in cMSSM+ due to the tweaking of its RG evolution by
the adjoint contribution.  In Fig.~\ref{fig4}, we choose the same high
scale parameters, except that now $A_0=[0,-4]$ TeV, so that cMSSM can
accommodate a 125 GeV mass of the Higgs.  The shaded regions in the
plane of the gluino and the lighter stop masses correspond to points
for which the light Higgs weighs between 123 and 127 GeV.  What we
demonstrate in Fig.~\ref{fig4} is that there is a significant recovery
of the lighter spectrum in the cMSSM+ compared to cMSSM.  We have made
use of two packages, \texttt{SuSpect} \cite{Djouadi:2002ze} and
\texttt{micrOMEGAs} \cite{Belanger:2010gh}, during the implementation
of these plots, and we have ensured that the predictions for some low
scale observables, e.g. $B_s$ decays to $\mu^+ \mu^-$, are consistent
with their experimental observations in the shaded regions.

We now give a quantitative estimate of how much we gain in terms of
fine-tuning. Using the Barbieri-Giudice criterion
\cite{Barbieri:1987fn}, a rather crude estimate of the amount of
fine-tuning as a function of the stop masses is given by
\cite{hep-ph/9801449}
\begin{equation}
\Delta \approx \frac{10}{33} \cdot \frac{m_{\tilde{t_1}} m_{\tilde{t_2}}
}{\left(650~{\rm GeV}\right)^2} \cdot \ln\left(\frac{M_G}{M_Z}\right) \, ,
\label{delta}  
\end{equation}
where it is to be noted that $\ln(M_G/M_Z) \simeq 33$.  Using a fixed
set of high scale input parameters, we present two sample points for
stop masses at the weak scale in GeV: $(m_{\tilde{t_1}},
m_{\tilde{t_2}}) =$ (1186, 1950) for cMSSM and (512, 902) for cMSSM+.
Both these points correspond to $m_h = 123$ GeV. Using
Eq.~(\ref{delta}), we obtain $\Delta \simeq 54$ for cMSSM and $\Delta
\simeq 11$ for cMSSM+. Thus we gain roughly a factor of 5 in terms of
fine-tuning, though this rough estimate which we presented for
illustration is only for a specific set of sample points.

We briefly mention some phenomenological implications of the color
adjoint superfield.  As shown in \cite{Cui:2006nw}, its scalar
component (assumed heavier) can decay into its fermionic component and
gluino much before the time of nucleosynthesis to avoid any
cosmological problems. The fermionic state can decay into
characteristic $4 t +$ neutralino missing energy through
nonrenormalizable interactions, which has very low standard
background. If the fermionic component is long-lived, there are ways
to ensure that its relic density is insignificant. In the LHC context
it has been shown that the experimental bounds on such models with
Dirac gauginos are weaker than on similar MSSM-type models
\cite{Kribs:2012gx}.  Several other authors have studied different
hadronic decay signatures of color-octet scalars and pseudo-scalars
\cite{Gerbush:2007fe,Schumann:2011ji,Zerwekh:2008mn}. The outcome of
our analysis in the context of the Higgs mass may provide further
motivation for the LHC studies of the adjoint states.

\noindent {\bf Conclusion and outlook:}~Na\"ively, one might think
that the presence of any colored matter would induce a similar shift
in the Higgs mass. Indeed so, but not always in the preferred
direction. We tried also with diquarks ($\Phi$) appearing through a
superpotential $W_{\rm DQ} = y_\Phi^{ij} u^c_i u^c_j \Phi + \mu_\Phi
\Phi \bar \Phi $, where for illustration we considered coupling with
up-type singlets. Here $\Phi = (\overline{3}, 1, 4/3)$ and $\bar \Phi
= (3, 1, -4/3)$ (see, e.g. Ref.~\cite{Bhattacharyya:1995bw} for a list
of possibilities for different diquark representations).  It is easy
to check that $\Delta b_3 = 1$ in this case, which is to be contrasted
with $\Delta b_3 = 3$ for the adjoint -- see Eq.~(\ref{b3}). This is
one of the reasons behind the much smaller shift in the Higgs mass
that one can get with a diquark.  Indeed in the diquark case, $\Delta
b_2$ and $\Delta b_1$ would be non-vanishing, but these are
numerically not so relevant in this context. Crucially, the diquark
Yukawa coupling $y_\Phi$ contributes in the `wrong' direction to the
running of $y_t$ and $A_t$, and its magnitude has to be kept under
control as otherwise $y_t$ would blow up pretty fast.  We relegate a
more detailed study of different types of diquarks in this context to
a future publication.  This last observation of ours, i.e. Yukawa
running in this context is indeed a tricky issue, is in accord with a
recent study claiming that additional chiral fermions at the GUT scale
with large Yukawa couplings modify $A_t$ in a way that the light Higgs
mass is actually {\em reduced} for the same stop and gluino masses
\cite{vempati}.  Herein lies the reason as to why the color adjoint
extension offers the most promising scenario in the present context.

We advance three distinct features which make our scenario a worthy
competitor of the alternative avenues for providing a few extra GeV to
the Higgs mass: ($i$) The existence of a chiral superfield in the
adjoint representation is well motivated as arising from higher
dimensional supersymmetric theory. ($ii$) The colored fermionic
component may provide a large Dirac mass of the gluino, which offers
many interesting features, including an improved naturalness.  We have
also shown using the Barbieri-Giudice criterion that fine-tuning in
cMSSM+ is lessened by a factor $\sim 5$ with respect to cMSSM
. ($iii$) Even if the adjoint scalar and fermion states are much
heavier than 1 TeV, e.g. if we assume them to weigh around 10 TeV, our
main conclusion remains unaffected. The central issue is what is the
value of the strong coupling when it stops running (at one loop
level). If the adjoint scalar and fermion masses are about 1 TeV, then
$\alpha_3^{-1} (1~{\rm TeV}) \simeq 9.7$ is where the strong coupling
freezes and stays unmoved for higher energies up to the GUT scale. On
the other hand, if the adjoint scalar and fermion masses are around 10
TeV, then the slope of the strong coupling keeps changing up to the
energy scale of 10 TeV, and then as soon as the adjoint states are
excited the coupling gets frozen at $\alpha_3^{-1} (10~{\rm TeV})
\simeq 10.8$, which is much closer to its value at 1 TeV and quite far
away from the GUT scale value (in cMSSM, i.e. without the adjoint
state) $\alpha_G^{-1} \simeq 23.4$.  As a result, even for a 10 TeV
adjoint state, a few GeV enhancement to the Higgs mass would not be a
problem. Indeed, the collider phenomenology of a 10 TeV state is less
exciting, which is why we assumed the adjoint states to weigh around 1
TeV. In this sense our scenario offers a rather robust mechanism for
the incremental shift in the Higgs mass.

We reiterate that by no means one can say that cMSSM (or,
equivalently, mSUGRA) is already disfavored. All we observe is that
the light Higgs mass in cMSSM struggles to reach out to the last few
rungs of its experimental range. This can be considerably eased if,
instead of holding ourselves hostage to the conventional particle
content of the MSSM, we add a new TeV scale color adjoint superfield,
which has enough motivations to exist and which offers rich
phenomenology to be explored at the LHC.

\noindent {\bf Acknowledgments:}~ G.B. acknowledges DFG support
through a Mercator visiting professorship, grant number INST
212/289-1, and hospitality at T.U. Dortmund. He also thanks CEA,
Saclay, for a short visit where the project was initiated.  The
research of T.S.R. is supported in part by the Agence Nationale de la
Recherche under contract ANR 2010 BLANC 041301, the European
Commission under Contract PITN-GA-2009-237920 (UNILHC) and the
Australian Research Council.  T.S.R. also acknowledges hospitality at
T.U. Dortmund and Department of Physics, Calcutta University, during
different stages of the work. We thank T.~Gherghetta for discussions.



\end{document}